\documentstyle{article}
\font\cero=cmss10 scaled 1728 \font\uno=cmssbx10 scaled 1200
\begin{document}
\begin{flushleft}
{\cero A dynamical metric and its ground state from the breaking down of the topological invariance of the Euler characteristic}\\
\end{flushleft}
{\sf R. Cartas-Fuentevilla, A. Escalante-Hernandez, A. Herrera-Aguilar}
  \\
{\it Instituto de F\'{\i}sica, Universidad Aut\'onoma de Puebla, Av. San Claudio y 18 Sur, Col. San Manuel, Ciudad Universitaria;
Apartado postal J-48 72570 Puebla Pue., M\'exico}; \\

\noindent{\sf R. Navarro-Perez} \\
{\it
Departamento de Actuaria, F\'{\i}sica y Matem\'{a}ticas, Universidad de las Am\'{e}ricas, Puebla,
Apartado postal 408, Ex-hacienda Sta. Catarina M\'{a}rtir, 72820 Cholula, Pue.,
M\'exico; and Departamento de F\'{\i}sica At\'{o}mica, Molecular y Nuclear, Facultad de Ciencias, Campus Fuentenueva S/N, Granada 18071, Espa\~{n}a} \\

Quantum  state wave functionals are constructed in exact form
for the graviton-like field theory obtained
by breaking down the topological symmetry of
the string action related with the Euler characteristic of the world-surface;
their continuous and discrete symmetries are discussed. The comparison with the so-called
Chern-Simons state, which may be inappropriate as quantum state, allows us to conclude that the found wave functionals will give a plausible approximation to the ground state for the considered field theory.\\


 \noindent {\uno I. Introduction} \vspace{1em}

One fundamental aspect in the study of dynamical systems and particularly of field theories is the finding of the ground state or vacuum, whose excitations will generate all other physically relevant states of the theory. In most of field theories, the ground state can be determined as the stable solution with the lowest energy; however, in those field theories involving the background independence symmetry like gravity this criterion is not physically admissible since the concept of energy is very subtle \cite{1a}. In the case of field theories with gauge symmetries, which correspond to first class constrained systems, a plausible representation of the vacuum state will be given by the wave functional solving all fist class constraints of the theory, and in the specific case of theories that involve gravity as an essential ingredient, the wave functional must solve particularly the vanishing extended Hamiltonian which corresponds to a linear combination of first class  constraints.

In the cases where an energy-momentum tensor $(T^{\mu\nu})$ can be constructed, a Hamiltonian $\cal H$ can be obtained through the identification ${\cal H} \equiv \int T^{00}$, and hence a possible representation $(\psi)$ of the vacuum state will satisfy the Schr\"{o}dinger equation $\widehat{\cal H}\psi = E\psi$, once $\cal H$ has been promoted to a quantum operator $\widehat{\cal H}$, and the positivity of the energy implies that the lowest-energy state will satisfy the vanishing Hamiltonian condition $\widehat{\cal H}\psi=0$; by this way the celebrated Chern-Simons state was constructed as an exact wave functional of the vacuum of Yang-Mills theory; this issue will be discussed in detail throughout the paper, since the wave functionals considered in the present treatment will be constructed using precisely the Schr\"{o}dinger representation, and a direct comparison with the Chern-Simons state is mandatory in order to establish similarities and differences, and subsequently to consider their possible viability as quantum states.

On the other hand, in a recent paper \cite{1} a spin-two field theory was constructed by
considering perturbative expansions around the topological sector
described by the number of handles of the string world-surface; these results are unexpected
and very surprising, considering that a topological field theory is devoid of physical degrees of freedom,
and in certain sense is trivial, however its symmetries are close to General Relativity symmetries \cite{2a}. In this manner, the results in \cite{1} show that perturbatively a topological field theory may have local degrees of freedom, in such a way that if all perturbative orders
are taken into account, the complete theory effectively has no local degrees of freedom, conciliating these new results with the usual ideas on a topological field theory. It is worth to comment that there is a similar scenario in gravity and linearized gravity; in fact, if a perturbation of the fundamental metric is considered around the fixed Minkowski spacetime, then the background independence symmetry is broken, and the image of a dynamical gravitational field is missed. Furthermore,  the extended Hamiltonian does not vanish and active diffeomorphisms are not the fundamental gauge symmetries  anymore. In this respect, gravity and linearized gravity are two different theories to each other. Hence, perturbative methods in general covariant theories could allow  emerging  theories with different symmetries and it is mandatory to study them.  As we shall see in the present paper, in topological theories  the perturbative expansion breaks down the topological symmetry of the original action, and a spin-two field emerges. The graviton-like field theory
obtained at first perturbative order in the world-surface metric, is a physically acceptable field theory: a) the corresponding action is quadratic in world-surface derivatives of the spin-two field; b) the equations of motion derived variationally from the action correspond to  wave-like equations describing propagating fields on the entire world-surface; c) the scalar curvature $R$ of the (unperturbed) world-surface plays the role of an {\it ``effective mass}" for the graviton-like field, in such a way that  massive and massless modes are present; d) there exists a correspondence between invariance of the action under diffeomorphisms (understandable as a gauge symmetry) and masslessness; in the massive mode the action contains a mass term that breaks down the gauge symmetry of the massless case; e) if the possible sources for the world-surface metric perturbations are considered, then
 a consistent conservation on both sides of the equality can be established: 
 
{\it deformations of the world-surface geometry = sources for the deformations}.

 Therefore there is nothing weird or exotic in this field theory, rather it looks like a conventional and consistent gauge theory that deserves a profound study; the present work represents a step in this direction. With this purpose, in the next section we outline the results obtained in \cite{1}, introducing definitions, conventions, and basic ideas. In Section III, the energy-momentum tensor for the spin-2 field is constructed considering the variations of the action with respect to the world-surface metric; the result is a symmetric tensor, from which a Hamiltonian can be chosen. In Section IV, the quantization of the theory in the Schr\"{o}dinger functional representation is considered; the wave functionals that satisfy the zero-energy condition for the Hamiltonian functional operator are constructed in an exact form. This requires self-dual, or anti-self-dual field configurations, leading to the concept of instantons, or anti-instantons for this graviton-like field, such as in a conventional gauge theory; the ground states are in the kernel of the operator versions of the self-dual, or anti-self-dual conditions. In Section V the continuous and discrete symmetries of the theory and its ground states are considered.
 In Section VI we develop the analysis of this theory as a dynamical phase for the world-sheet metric in contrast to the topological phase of the original Euler characteristic. Additionally in Section VII the analysis of the waves propagation obtained from
 the quadratic action by expanding the inner curvature around a fixed world-surface metric is developed in order to contrast with those obtained
 throughout the paper. In Sections VIII and IX we consider explicit background world-surface scenarios, $S^{2}$, and $T^{2}$.
  We finish in Section X with discussions and perspectives.\\

\noindent {\uno II. Preliminaries of the world-surface graviton
field theory}
\vspace{1em}

We outline the description given in \cite{2} for
 a space-like or
time-like $2$-surface imbedded in an $n$-dimensional space or
space-time background with metric $g_{\mu\nu}$, which can be decomposed as:
\[
     n^{\mu} {_{\nu}} + \bot^{\mu} {_{\nu}} = g^{\mu} {_{\nu}},
     \quad n^{\mu} {_{\nu}} \ \bot^{\nu} {_{\rho}} = 0,
\]
where $n^{\mu\nu}$ is the (first) {\it fundamental tensor} of the
$2$-surface, and $\bot^{\mu\nu}$ the complementary orthogonal
projection. The tangential covariant differentiation operator is defined
in terms of the fundamental tensor as
\[
     \overline{\nabla}_{\mu} = n^{\rho} {_{\mu}} \ \nabla_{\rho},
\]
where $\nabla_{\rho}$ is the usual Riemannian covariant
differentiation operator associated with $g_{\mu\nu}$.
Additionally, the second fundamental tensor is defined by
\[
     K_{\lambda\mu}{^{\nu}} = n^{\sigma}_{\mu} \
     \overline{\nabla}_{\lambda} \ n^{\nu}_{\sigma} =
     K_{(\lambda\mu)}{^{\nu}},
\]
with its property of tangentiality of the first two indices and
orthogonality of the last one.

In \cite{1}, the adjusted tangential covariant
differentiation $\widetilde{\nabla}_{\mu}$  in terms of
$\overline{\nabla}_{\mu}$ and $K_{\mu\nu}{^{\lambda}}$ was introduced,
\[
     \widetilde{\nabla}_{\mu} A_{\alpha\beta} \equiv \overline{\nabla}_{\mu}
     A_{\alpha\beta} - K_{\mu}{^{\rho}}{_{\alpha}} A_{\rho\beta} -
     K_{\mu}{^{\rho}}{_{\beta}} A_{\alpha\rho},
\]
where $A_{\alpha\beta}$ is an arbitrary (world-sheet tangent)
field. For example, it is easy to prove using the two previous definitions
that
\[
     \widetilde{\nabla}_{\mu} n_{\alpha\beta} = 0.
\]
In this manner, $\widetilde{{\nabla}}$ will operate as a
``connection compatible" with the fundamental tensor
$n_{\alpha\beta}$ of the world-surface. Additionally we can show that \cite{1}
\[
     (\widetilde{\nabla}_{\alpha} \widetilde{\nabla}_{\sigma} -
     \widetilde{\nabla}_{\sigma} \widetilde{\nabla}_{\alpha})
     A_{\mu\nu} = R_{\alpha\sigma\mu} {^{\rho}} A_{\rho\nu} +
     R_{\alpha\sigma\nu} {^{\rho}} A_{\mu\rho},
\]
however, in the case of a 2-surface,  the internal curvature of the world-surface $R_{\kappa\lambda}{^{\mu\nu}}= R
n^{[\mu}_{[\kappa}n^{\nu]}_{\lambda]}$ \cite{2}, and invoking the conformal symmetry of the world-surface,
we can fix (the {\it effective mass} for the graviton-like field \cite{1}) $R=0$, and consequently from the above equation,
\[
     \widetilde{\nabla}_{\alpha} \widetilde{\nabla}_{\sigma} -
     \widetilde{\nabla}_{\sigma} \widetilde{\nabla}_{\alpha}
      =0;
\]
which we shall use implicitly in our calculations. In general the commutator of background covariant derivatives ${\nabla}$, and of their world-surface projections $\overline{\nabla}$ does not vanish. We can form vanishing-commutator operators only in the combination given by $\widetilde{\nabla}$; this is possible due to the integrability condition involving the world-surface curvature, the second fundamental tensor, and world-surface projections of the background curvature \cite{2}.

In \cite{1} the following action was constructed for a world-surface
graviton field,
\begin{equation}
     S =\frac{1}{2} \int d\overline{\Sigma} (-\widetilde{\nabla}_{\sigma}
     H_{\alpha\beta} \widetilde{\nabla}^{\sigma} H^{\alpha\beta} +
     2 \widetilde{\nabla}^{\alpha} H_{\alpha\beta}
     \widetilde{\nabla}_{\sigma} H^{\sigma\beta} -2
     \widetilde{\nabla}^{\sigma} H \widetilde{\nabla}^{\rho}
     H_{\rho\sigma} - k H_{\alpha\beta}
     \overline{S}^{\alpha\beta}),
\end{equation}
where $H^{\mu\nu}$ represents the trace-free part of the infinitesimal fluctuations of the first fundamental tensor, $H^{\mu\nu} \equiv \delta n ^{\mu\nu}- \frac{1}{2} n ^{\mu\nu}( n_{\alpha\beta}\delta n ^{\alpha\beta})$, and the integration is on the world-surface characterized topologically by its Euler characteristic. This action
is invariant under the action of diffeomorphisms in the case
when the world-surface scalar curvature $R$  vanishes \cite{1}. Although $H=n_{\alpha\beta}H^{\alpha\beta}=0$, it is convenient to include a term involving $H$ in the action. In fact, just like in linearized gravity first are considered all variations involving the action, then the gauge fixing  is imposed  on the spin-two field \cite{3a}.    The last term of the form $k H_{\alpha\beta}
     \overline{S}^{\alpha\beta}$ represents an interacting term of $H_{\alpha\beta}$ with its possible sources $\overline{S}^{\alpha\beta}$, $k$ is the coupling constant. We consider the source-free case, {\it i.e.} $\overline{S}^{\alpha\beta}=0$.

Variationally we can obtain from the action (1) the corresponding
equations of motion and additionally to identify the phase space
canonical variables in order to quantize the theory:
\begin{eqnarray}
     \delta S \!\! & = & \!\! \int d\overline{\Sigma} [
     \widetilde{\Box} H^{\alpha\beta} - \widetilde{\nabla}^{\alpha}
     \widetilde{\nabla}_{\sigma} H^{\sigma\beta} - \widetilde{\nabla}^{\beta}
     \widetilde{\nabla}_{\sigma} H^{\sigma\alpha} + n^{\alpha\beta}
     \widetilde{\nabla}^{\rho} \widetilde{\nabla}^{\sigma}
     H_{\rho\sigma}] \delta H_{\alpha\beta} \nonumber \\
     \!\! & & \!\! + \int d \overline{\Sigma}
     \overline{\nabla}_{\sigma} [ (2 n^{\sigma (\alpha}
     \widetilde{\nabla}_{\rho} H^{\beta )\rho} -
     \widetilde{\nabla}^{\sigma} H^{\alpha\beta} - n^{\alpha\beta}
     \widetilde{\nabla}_{\rho}  H^{\rho\sigma}) \delta
     H_{\alpha\beta}];
     \nonumber
\end{eqnarray}
from the first term the equations of the motion read
\begin{equation}
     \widetilde{\Box} H^{\alpha\beta} - \widetilde{\nabla}^{\alpha}
     \widetilde{\nabla}_{\sigma} H^{\sigma\beta} -
     \widetilde{\nabla}^{\beta} \widetilde{\nabla}_{\sigma} H^{\sigma\alpha}
     + n^{\alpha\beta} \widetilde{\nabla}^{\rho} \widetilde{\nabla}^{\sigma}
     H_{\rho\sigma} = 0;
     \label{eq-motion}
\end{equation}
the second term is understandable as the divergence of a {\it phase space
symplectic potential} of the form $``p\delta q"$, from which a canonical pair $(p,q)$ is
recognized:
\begin{equation}
     (P^{\sigma\alpha\beta}, H_{\alpha\beta}) \equiv (
     \widetilde{\nabla}^{\sigma} H^{\alpha\beta} + n^{\alpha\beta}
     \widetilde{\nabla}_{\rho}  H^{\rho\sigma} - 2 n^{\sigma
     (\alpha} \widetilde{\nabla}_{\rho} H^{\beta )\rho},
     H_{\alpha\beta});
\end{equation}
note that $(P^{\sigma\alpha\beta}, H_{\alpha\beta})$ are world-surface tangent in all indices;
moreover, the conjugate momentum $P$ has the following properties
with respect to the last indices:
\begin{itemize}
     \item[a)]  symmetric: $P^{\sigma\alpha\beta} =
     P^{\sigma\beta\alpha}$;
     \item[b)]  traceless: $n_{\alpha\beta} P^{\sigma\alpha\beta} = 0$;
\end{itemize}
and is conserved in relation to the first index
\begin{itemize}     
	\item[c)]  
     $\widetilde{\nabla}_{\sigma} P^{\sigma\alpha\beta} = 0$
\end{itemize}     
as a consequence of equations of motion (2); in fact, the equations of motion take this form of conservation of the momentum; and finally
\begin{itemize}  
     \item[d)]  $n_{\sigma\alpha} P^{\sigma\alpha\beta} = -
     \widetilde{\nabla}_{\sigma} H^{\sigma\beta}$, 
\end{itemize}
which will vanish under the gauge fixing condition considered bellow.

In order to simplify the treatment it will be convenient to invoke
the invariance under diffeomorphisms proved in \cite{1}, and to
impose the {\it gauge condition}
\begin{equation}
     \widetilde{\nabla}_{\sigma} H^{\sigma\beta} = 0,
\end{equation}
obtaining the following simplifications:
\begin{itemize}
  \item[A)]  $P^{\sigma\alpha\beta} = \widetilde{\nabla}^{\sigma}
  H^{\alpha\beta}, \quad \quad n_{\sigma\alpha} P^{\sigma\alpha\beta} = 0$;
  \item[B)]  the equations of motion (2) reduce to: $\widetilde{\Box}
  H^{\alpha\beta} = \widetilde{\nabla}_{\sigma} P^{\sigma\alpha\beta} = 0$;
  \item[C)] the action (1) reduces to a quadratic functional in the
  momentum: $S = -\frac{1}{2} \int d\overline{\Sigma}
  P^{\sigma\alpha\beta} P_{\sigma\alpha\beta}$; this property will be very convenient in the path integral formulation of the theory (see Section 9 in \cite{1});
  \item[D)] as mentioned in the introduction, a consistent conservation law
            between metric deformations of the world-surface and their possible sources can be established from the first-order deformation of the Bianchi identities [1]; specifically the identity
            $
     \overline{\nabla}_{\alpha}(\widetilde{\Box}
  H^{\alpha\beta}) =
     K_{\alpha\sigma}{^{\beta}}(\widetilde{\Box}
  H^{\alpha\sigma})$, was obtained; this identity guarantees the $\widetilde{\nabla}$-conservation of both members of equations of motion (2) when the possible sources
  $\overline{S}^{\alpha\beta}$ are included on the left hand side (for more details see Section 5 in [1]).
\end{itemize}

\noindent {\uno III. The energy-momentum tensor for the
graviton-like field}
\vspace{1em}

We define the energy-momentum tensor for the field as usual, the
functional derivative of the action (1) with respect to the {\it
world-surface metric} $n^{\mu\nu}$: $ T_{\mu\nu} = \frac{\delta S}{\delta
n^{\mu\nu}}$. Thus, it is convenient to rewrite the Lagrangian in
a $n^{\mu\nu}$-explicit dependent way:
\[
     L = \frac{1}{2} (2 n^{\alpha\sigma} n^{\lambda\rho} - n^{\sigma\lambda}
     n^{\alpha\rho} ) n^{\beta\gamma} \overline{\nabla}_{\sigma}
     H_{\alpha\beta}  \overline{\nabla}_{\lambda} H_{\rho\gamma};
\]
a direct calculation gives
\begin{equation}
     T_{\mu\nu} = \frac{\delta S}{\delta n^{\mu\nu}} = \{
     n^{\beta\gamma} [ n^{\lambda\rho} \delta^{\alpha}_{\mu}
     \delta^{\sigma}_{\nu}  + n^{\alpha\sigma} \delta^{\lambda}_{\mu}
     \delta^{\rho}_{\nu} - n^{\sigma\lambda} \delta^{\alpha}_{\mu}
     \delta^{\rho}_{\nu} - \frac{1}{2} n^{\alpha\rho}
     \delta^{\sigma}_{\mu} \delta^{\lambda}_{\nu}] + n^{\sigma\alpha}
     n^{\lambda\rho} \delta^{\gamma}_{\mu} \delta^{\beta}_{\nu} \}
     \overline{\nabla}_{\sigma} H_{\alpha\beta}
     \overline{\nabla}_{\lambda} H_{\rho\gamma} - \frac{1}{2}
     n_{\mu\nu} L;
\end{equation}
where the last term proportional to $L$  comes from the variation of
the world-surface element: $\delta d \overline{\Sigma} = -
\frac{1}{2} n_{\mu\nu} \delta n^{\mu\nu} d \overline{\Sigma}$. Note
that $T^{\mu\nu}$ is symmetric, with a non-vanishing trace, $ n^{\mu\nu}
T_{\mu\nu} = 2L$.

The general expression (5) is very difficult to work out; however
with the gauge fixing condition (4) and the subsequent
simplifications A), B), C) considered in Section II, it takes a particularly
simple form:
\begin{equation}
     T^{\mu\nu} = \frac{1}{2} [ \frac{1}{2} n^{\mu\nu}
     P^{\lambda\rho\gamma} P_{\lambda\rho\gamma} - P^{\mu}
     {_{\lambda\rho}} P^{\nu\lambda\rho} - 2 P^{\lambda\rho\mu}
     P_{\lambda\rho} {^{\nu}} ];
\end{equation}
from which we can define a Hamiltonian $H$ for the theory (the
factor 4 is for future convenience),
\begin{equation}
     H = 4 T^{00} = 2 \big[ \frac{1}{2} n^{00} P^{\lambda\rho\gamma}
     P_{\lambda\rho\gamma} - P^{0} {_{\lambda\rho}} P^{0\lambda\rho}
     - 2 P^{\lambda\rho 0} P_{\lambda\rho} {^{0}} \big].
\end{equation}
Further simplifications can be obtained considering that the gauge condition
(4) represents $n$ conditions  on the field $H_{\mu\nu}$, and then
we can choose $H^{0\alpha} = 0, \alpha = 0, \ldots, n-1$, this conditions are identified as Lagrange multipliers in a conventional canonical scheme,  leaving
$H^{ij} \neq 0$ with $i,j = 1,...,n-1$; symbolically $ (H^{\mu\nu}) =
\pmatrix{0 \quad 0 \cr 0 \quad H_{ij} \cr}$. Consequently from the
definition (3) we have for the momentum $P^{\sigma 0\alpha} = 0$, leaving $P^{0ij}
\neq 0$, and $P^{ijk} \neq 0$; note that in particular the conjugate
momenta to the Lagrange multipliers  $H^{0\alpha}$ vanish. With these considerations we have
that
\begin{equation}
     H = 2 \big[ \frac{1}{2} n^{00} P^{0ij} P_{0ij} + \frac{1}{2}
     n^{00} P^{ijk} P_{ijk} - P^{0}{_{ij}} P^{0ij} \big] = - P^{0ij}
     P^{0}{_{ij}} + n^{00} P^{ijk} P_{ijk}.
\end{equation}
We need to decompose the ambient space $M$ in order to develop the quantization.
The symplectic potential gives a possible
decomposition; explicitly it
reads $\theta = \int_{\Sigma^{C}} P^{\sigma\alpha\beta} \delta
H_{\alpha\beta} d \Sigma^{C}_{\sigma}$, with $M = \Sigma^{C} \times R$, where
$\Sigma^{C}$ corresponds to a Cauchy space-like surface (which contains the {\it spatial} configuration of the string), and $R$ may be a
time-like direction; thus $\theta$ takes the form $\theta =
\int_{\Sigma^{C}} P^{0\alpha\beta} \delta H_{\alpha\beta} d
\Sigma^{C}_{0}$, and the canonical pair given by $(P^{0\alpha\beta},
H_{\alpha\beta})$, reduces to $(P^{0ij}, H_{ij})$, considering the above restrictions. In this manner the components
$P^{0ij}$ are considered as phase space functionals of $H_{ij}$. At
this point it is convenient to establish an ``electro-magnetic"
analogy, which allows us to give a parallel treatment.

Thus, we define the ``electric" components of $P$ as $P^{0ij} \equiv
E^{ij}$, and the magnetic part as $ P^{ijk} \equiv B^{ijk}$. For
example, in the QED case the canonical pair is given by $( E^{i},
A_{i})$, and the magnetic field is considered as a functional of $A_{i}$,
$B^{i} = \epsilon^{ijk} F_{jk} (A)$.

Hence, the Hamiltonian (8) can be rewritten as
\begin{equation}
     H = - E^{ij} E_{ij} + n^{00} B^{ijk} B_{ijk}.
\end{equation}
Furthermore, the {\it magnetic part} of the expression (9) can be rewritten in the following
way, by considering the above space-time decomposition, and the relation $n^{\mu}{_{\nu}} = \mp
\varepsilon_{\sigma}{^{\mu}} \varepsilon^{\sigma}{_{\nu}}$ [where
the upper (lower) sign applies to the case of a time-like
(space-like) world-sheet \cite{2}], which reduces in particular to $n^{k}{_{l}} = \mp
\varepsilon_{0}{^{k}} \varepsilon^{0}{_{l}}$:
\[
    (n^{00}) B^{ijk} B_{ijk} =(n^{00})  n^{k}{_{l}} B^{ijl} B_{ijk} = \mp
     (n^{00}) (\varepsilon^{0}{_{l}} B^{ijl}) (\varepsilon_{0}{^{k}} B_{ijk})
     = \mp {^{\ast}} B ^{ij} {^{\ast}} B_{ij},
\]
where we define the (world-surface) dual to the {\it magnetic field}
$\ {^{\ast}}B^{ij} \equiv \varepsilon^{0}{_{l}} B^{ijl}$. Therefore the
Hamiltonian (9) can be rewritten in the following {\it factorized}
form:
\begin{equation}
    H = - E^{ij} E_{ij} \mp {^{\ast}} B^{ij} {^{\ast}} B_{ij} =
    \left\{ \begin{array}{ll}
    -(E^{ij} + i {^{\ast}} B^{ij}) (E_{ij} - i {^{\ast}} B_{ij}), &
    \rm{time-like \quad world-sheet}; \\
    -(E^{ij} - {^{\ast}} B^{ij}) (E_{ij} + {^{\ast}} B_{ij}), &
    \rm{space-like \quad world-sheet};
    \end{array} \right.
\end{equation}
The factorized form will be crucial in the reduction of the zero-energy condition on the Hamiltonian operator to the quantum self-dual conditions, whose integration is simpler.\\

\noindent {\uno IV. Quantization and graviton-like instantons}
\vspace{1em}

In order to give a functional Schr\"{o}dinger
representation of the theory, we use the polarization in terms of
the canonical phase space variables $(E^{ij}, H_{ij})$. In this
representation a state $|\psi
>$ will be represented by a functional depending on
$H_{ij}$, which will operate by a functional multiplication, and
the canonical momentum $E^{ij}$ by (complex) functional differentiation with
respect to $H_{ij}$:
\begin{eqnarray}
     |\psi > \!\! & \rightarrow & \!\! \psi (H_{ij}), \nonumber \\
     H_{ij} |\psi > \!\! & \rightarrow & \!\! H_{ij} \psi (H_{lm}),
     \nonumber \\
      E^{ij} |\psi > \!\! & \rightarrow & \!\! i \frac{\delta}{\delta
     H_{ij}} \psi (H_{lm}), \nonumber \\
     B^{ijk} |\psi > \!\! & \rightarrow & \!\! B^{ijk} \psi
     (H_{lm}), \\
     {^{\ast}} B ^{ij} |\psi > \!\! & \rightarrow & \!\! {^{\ast}} B^{ij}
     \psi (H_{lm}); \nonumber
\end{eqnarray}
with this classical-quantum correspondence, we develop explicitly the quantization for the
space-like world-surface case; the time-like case has an entirely similar treatment.\\

\noindent {\uno IVa. A space-like world-surface}
\vspace{1em}

Considering the expressions (10) and (11), the quantum Hamiltonian operator will take the form
\begin{equation}
     H = - \big( i \frac{\delta}{\delta H_{ij}} + {^{\ast}} B^{ij}
     \big) \big( i \frac{\delta}{\delta H^{ij}} + {^{\ast}} B_{ij}
     \big).
\end{equation}
A representation of the vacuum state for the world-surface graviton-like field is given for a ground state wave functional $\psi$
satisfying the zero-energy condition on the Hamiltonian operator:
\begin{equation}
     H\psi = - \big( i \frac{\delta}{\delta H_{ij}} + {^{\ast}} B^{ij}
     \big) \big( i \frac{\delta}{\delta H^{ij}} + {^{\ast}} B_{ij}
     \big) \psi = 0;
\end{equation}
the second-order functional derivative equation (13) can be reduced
fortunately to a first order functional derivative equation, since
any wave functional $\psi_{0}$ satisfying
\begin{equation}
     \big( i \frac{\delta}{\delta H_{ij}} \pm {^{\ast}} B^{ij}
     \big) \psi_{0} = 0,
\end{equation}
satisfies automatically (13). It is very easy to integrate Eq.\ (14), the solutions are given by phases,
\begin{equation}
     \psi_{0} = e^{\pm i \int_{\Sigma^{C}} {^{\ast}} B^{ij}H_{ij} d
     \Sigma^{C}_{0}}.
\end{equation}
Note that at classical level, the quantum conditions (14) correspond
to the conditions of self-duality, or anti-self-duality,
\begin{equation}
     E^{ij} = \pm {^{\ast}}B^{ij};
\end{equation}
in analogy to conventional gauge theories, we can define the (world-surface) graviton-like instantons as
the field configurations satisfying (16); thus, the ground states (15)
corresponds classically to instantons or anti-instantons. It is important
to mention that the conditions (16) are imposed on the fields associated with $H_{ij}$
(the metric fluctuations of the world-surface), and not on the geometry of the un-perturbed world-surface, {\it i.e.}
they represent no self-dual nor anti-self-dual world-surface manifolds.
\\

\noindent {\uno IVb. A time-like world-surface}
\vspace{1em}

Considering now the first of Eqs.\ (10), the Hamiltonian operator will take the form
\begin{equation}
     H = \big( \frac{\delta}{\delta H_{ij}} + {^{\ast}} B^{ij}
     \big) \big( \frac{\delta}{\delta H^{ij}} - {^{\ast}} B_{ij}
     \big);
\end{equation}
with all the $i$'s removed in relation to the previous case. The ground state functionals
satisfy now,
\begin{equation}
     \big( \frac{\delta}{\delta H_{ij}} \pm {^{\ast}} B^{ij}
     \big) \phi_{0} = 0,
\end{equation}
 with real solutions,
\begin{equation}
    \phi_{0} = e^{\pm \int_{\Sigma^{C}} {^{\ast}} B^{ij}H_{ij} d
     \Sigma^{C}_{0}}.
\end{equation}
 Classically, the quantum conditions (18) correspond
to the conditions of self-duality, or anti-self-duality:
\begin{equation}
     E^{ij} = \pm i{^{\ast}}B^{ij}.
\end{equation}
In the context of loop quantum gravity, the Lorentzian state does not contain an $i$
in the exponential of the Chern-Simons form, whereas the Euclideanized version is a phase, because
there is an $i$ in the exponential \cite{3}; in this sense the analogy to the present case is fulfilled.
 Particularly in the case of the real solutions (19), the Born-Oppenheimer method is in principle applicable to develop a {\it semiclassical} expansion around the ground state. This will be the subject of subsequent works.
\vspace{1em}

\noindent {\uno V. Symmetries of the theory and its ground states}
\vspace{1em}

The general covariance can be retrieved partially in the expression
for the argument of the wave functionals (15), and (19) as follows:
\begin{equation}
     {^{\ast}}B^{ij} H_{ij} d \Sigma^{C}_{0} = \varepsilon
     ^{0}{_{k}} B^{ijk} H_{ij} d \Sigma^{C}_{0} \rightarrow
     \varepsilon^{\mu} {_{\nu}} P^{\alpha\beta\nu} H_{\alpha\beta} d
     \Sigma^{C}_{\mu};
\end{equation}
similarly the conditions (16), and (20)  can be rewritten in the
covariant form
\begin{equation}
    P^{\mu\alpha\beta}  =
    \left\{ \begin{array}{ll}
    \pm i\varepsilon^{\mu} {_{\nu}}
     P^{\alpha\beta\nu} , &
    \rm{time-like \quad world-sheet}; \\
    \pm \varepsilon^{\mu} {_{\nu}}
     P^{\alpha\beta\nu}, &
    \rm{space-like \quad world-sheet};
    \end{array} \right.
 \end{equation}
The complete symmetry may be retrieved once we remove the
constraint (4); however, we restrict ourselves to the constraint formulation of
the theory for simplicity.

The covariant expressions (21), and (22) are consistent with the relationship between the actions
of different phases of the theory under the self-dual, or anti-self-dual conditions considered in detail in Section VI.\\

\noindent {\uno Va. Invariance under diffeomorphisms}
\vspace{1em}

Under the action of diffeomorphisms generated by an arbitrary field
$V^{\mu}$, the fields associated  with the world-surface graviton
undergo the changes \cite{1},
\begin{eqnarray}
     H'_{\mu\nu} \!\! & = & \!\! H_{\mu\nu} - \frac{1}{2}
     (\widetilde{\nabla}_{\mu} \overline{V}_{\nu} + \widetilde{\nabla}_{\nu}
     \overline{V}_{\mu} - n_{\mu\nu} \overline{\nabla}_{\alpha}
     \overline{V}^{\alpha}), \nonumber\\
     \widetilde{\nabla}^{\mu} H'_{\mu\nu} \!\! & = & \!\!
    \widetilde{\nabla}^{\mu} H_{\mu\nu} - \frac{1}{2}
     \widetilde{\Box} \overline{V}_{\nu},
\end{eqnarray}
where $\overline{V}^{\mu}
= n^{\mu}{_{\alpha}} V^{\alpha}$, {\it i.e.} a world-surface
diffeomorphism.

The action (1) turns out to be invariant under diffeomorphisms
\cite{1}; invoking this symmetry the constraint
$\widetilde{\nabla}_{\mu} H^{\mu\nu} = 0$ was imposed in order to
simplify the finding of the ground wave functionals; this invariance property was established as a gauge fixing  in \cite{1}. In fact, in the Hamiltonian context, this gauge fixing is equivalent to the Coulomb gauge used for instance in linearized gravity.  Hence, from the last of Eqs.\ (23) it follows that the constraint is possible provided that the vector
field $\overline{V}_{\nu}$ generating the diffeormophisms satisfies
\[
     \widetilde{\Box} \overline{V}_{\nu} = 0;
\]
if the ground states obtained within this constraint formulation
preserve that symmetry, then they must be invariant under the change
(23) modulo the constraint (and total derivatives); explicitly  we have for the argument of the wave functionals (21):
\begin{equation}
     P'^{\alpha\beta\sigma} H'_{\alpha\beta} = P^{\alpha\beta\sigma}
     H_{\alpha\beta} +\frac{\overline{V}_{\beta}}{2}\widetilde{\nabla}_{\alpha}P^{\alpha\beta\sigma},
\end{equation}
where the second term on the r.h.s. is proportional to the equations of motion (2) (see b) in Section II). In these
considerations the property $\widetilde{\nabla}_{\alpha} \varepsilon^{\mu}
{_{\nu}} = 0$ is taken into account, which is consequence of the
formula $\overline{\nabla}_{\sigma} \varepsilon^{\mu\nu} = 2
K_{\sigma\tau} {^{[\nu}} \varepsilon^{\mu ]\tau}$ \cite{2}.

In order to put in perspective our results, let us make a comparison
with the Yang-Mills case: although the four-dimensional Yang-Mills
action is gauge-invariant in the {\it exact} sense (without boundary
terms) and the gauge generator is the so-called Gauss constraint, its ground state is represented by the Chern-Simons wave
functional \cite{4}, which is gauge-invariant modulo a boundary term, under infinitesimal gauge transformations;
in the present case we have a  similar  situation, the original action
shows the symmetry in the {\it exact} sense,  the gauge generator is the Gauss-like constraint  $\tilde \nabla_i P^{0il}=0$, whereas the ground wave
functionals preserve that symmetry modulo a boundary term.\\

\noindent {\uno Vb. discrete symmetries: $\cal C,P,T$.}
\vspace{1em}

Since the action of the parity operator depends sensitively on the
dimension of the (ambient) space-time, we consider for simplicity
only the physically more interesting case, a four-dimensional
Minkowskian background.

In this case, the ambient orthonormal frame vectors consist of an
internal subset of vectors $\{ i^{\mu}_{0}, i^{\mu}_{1} \}$, where
$i^{\mu}_{0}$ is a time-like unit vector, and $i^{\mu}_{1}$, a
space-like unit vector, both tangent to the world-sheet; and an
external subject of space-like vectors $\{ \lambda^{\mu}_{1},
\lambda^{\mu}_{2} \}$, orthogonal to the world-sheet. Therefore the
parity operation $\cal P$ can be defined in terms of the space-like
vectors $\{ i^{\mu}_{1}, \lambda^{\mu}_{1}, \lambda^{\mu}_{2} \}$ as
\begin{equation}
     \{ i^{\mu}_{1}, \lambda^{\mu}_{1}, \lambda^{\mu}_{2} \}
     \stackrel{\cal P}{\longrightarrow} \{ -i^{\mu}_{1},
     -\lambda^{\mu}_{1}, -\lambda^{\mu}_{2} \}, \{ -i^{\mu}_{1},
     \lambda^{\mu}_{1}, \lambda^{\mu}_{2} \}, \{ i^{\mu}_{1},
     -\lambda^{\mu}_{1}, \lambda^{\mu}_{2} \}, \{ i^{\mu}_{1},
     \lambda^{\mu}_{1}, -\lambda^{\mu}_{2} \},
\end{equation}
every case with a negative determinant; we shall consider the first
case. Similarly, the {\it time reversal} transformation ${\cal T}$ is
 defined by
\begin{equation}
     i^{\mu}_{0} \stackrel{\cal T}{\longrightarrow} -i^{\mu}_{0};
\end{equation}
considering that the background Minkowskian metric has the form $(1,
-1, -1, -1)$, it is very easy to show that under the combined $\cal PT$
transformations, the ambient frame vectors
\begin{equation}
     \{ i^{\mu}_{A},\lambda^{\mu}_{R} \}
     \stackrel{\cal PT}{\longrightarrow} \{ -i^{\mu}_{A},
     -\lambda^{\mu}_{R} \};
\end{equation}
furthermore, since the ambient gradient $\partial_{\mu}$ can be
decomposed in terms of $\{ i^{\mu}_{A}, \lambda^{\mu}_{R} \}$, we
have
\begin{equation}
      \partial_{\mu} \stackrel{\cal PT}{\longrightarrow} - \partial_{\mu};
\end{equation}
the following tensors are invariant under $\cal PT$, as a consequence of
Eq.\ (27) and the definitions :
\begin{equation}
     ( \eta^{\mu\nu}, \varepsilon^{\mu\nu}, H^{\mu\nu}= \delta\eta^{\mu\nu}
     ) \stackrel{\cal PT}{\longrightarrow} (\eta^{\mu\nu}, \varepsilon^{\mu\nu},
     H^{\mu\nu} );
\end{equation}
on the other hand, the following composed tensors and operators change their
sign as a consequence of Eq.\ (28), and (29):
\begin{equation}
     (\overline{\partial}_{\mu} = \eta^{\nu}_{\mu} \partial_{\nu},
     K_{\mu\nu}{^{\lambda}}, \widetilde{\partial}_{\mu},
     P^{\alpha\mu\nu}, d\Sigma^{C}_{\mu})
     \stackrel{\cal PT}{\longrightarrow} (-\overline{\partial}_{\mu},
     -K_{\mu\nu}{^{\lambda}}, -\widetilde{\partial}_{\mu},
     -P^{\alpha\mu\nu}, -d\Sigma^{C}_{\mu}).
\end{equation}
As a consequence of these transformations we can establish the
discrete symmetries of the original action (1), and the ground state
wave functional (19). The action is quadratic in the
$\widetilde{\partial}_{\mu}$-derivatives of $H^{\mu\nu}$, and then
is trivially invariant under $\cal PT$. This symmetry is directly extended
to the equations of motion (2) (contain quadratic $\widetilde{\partial}$-derivatives of $H^{\mu\nu}$), the symplectic potential (contains two odd objects),
the energy-momentum tensor (6) (is quadratic
in $P^{\alpha\mu\nu}$), and in particular to the Hamiltonian (7). Furthermore, all
these geometric objects will be invariant under charge-conjugation
 ${\cal C}$, and hence we have a $\cal CPT$ symmetry at classical level.
 In the case of the ground state wave functional,
its integrant (21) transforms as
\begin{equation}
     \varepsilon^{\mu}{_{\nu}} P^{\alpha\beta\nu} H_{\alpha\beta}
     d\Sigma^{c}_{\mu} = \varepsilon^{\mu}{_{\nu}}
     \widetilde{\partial}^{\alpha} H^{\beta\nu} \cdot
     H_{\alpha\beta} d\Sigma^{C}_{\mu}
     \stackrel{\cal PT}{\longrightarrow} \varepsilon^{\mu}{_{\nu}}
     (-\widetilde{\partial}^{\alpha} H^{\beta\nu}) H_{\alpha\beta}
     (-d\Sigma^{c}_{\mu}) = \varepsilon^{\mu}{_{\nu}} P^{\alpha\beta\nu}
     H_{\alpha\beta} d\Sigma^{c}_{\mu},
\end{equation}
and hence, the quantum ground state is $\cal PT$-invariant, keeping the
symmetry of the classical field theory. $\cal T$ includes in general complex conjugation, being an
{\it anti-unitary} transformation; however, there is no $i$ in the argument (31) in the case of a
time-like world-surface, and the effect of a complex conjugation is imperceptible. On the other hand,
the self-duality, and anti-self-duality conditions (22) contain an $i$ in this case, and then that effect
is perceptible, in such a way that the transformation $\cal PT$ maps instantons into anti-instantons and {\it vice versa}.

In the case of a possible interacting term in the action of the form
$H_{\mu\nu} S^{\mu\nu}$, where $S^{\mu\nu}$ corresponds to the
sources of the world-surface metric deformations $H_{\mu\nu}$, one
expects $S^{\mu\nu}$ to be $\cal C$-invariant, since it is assumed that
particles and anti-particles generate the same geometric effects; in
this manner we have that the term $H_{\mu\nu} S^{\mu\nu}$ will be a
$\cal C$-invariant. Since $H_{\mu\nu}$ is separately $\cal CPT$-invariant, then
$S^{\mu\nu}$ must be a $\cal PT$ -invariant, in order to have a
$\cal CPT$-invariant interacting term.

As a partial conclusion of this Section we can say that the constructed functionals correspond to an admissible quantum state since they show invariance under $\cal CPT$  transformations, and this fact establishes a difference to favor these functionals in relation to the usual Chern-Simons state which is not $\cal CPT$-invariant, and consequently inadmissible as a quantum state \cite{5}.\\

\noindent {\uno VI. Topological and dynamical phases}
\vspace{1em}

In the context of four-dimensional Yang-Mills theory, it is well known that the self-dual conditions on the curvature deform the original action into a topological term, specifically the second Chern class of the associated bundles, which can be expressed as the (exterior) derivative of the corresponding Chern-Simons form; the bound state is given precisely by the exponential of such a Chern-Simons form, symbolically
\begin{equation}
\int(F\wedge \ast F)\quad \overrightarrow{_{\cal F=\pm \ast F}}\int(F\wedge F) =
\int d (\rm {Chern-Simons})+\int A\wedge DF,
\quad \Psi_{\rm {bound-state}}= e^{\int\rm {Chern-Simons}};
\end{equation}
however $DF=0$, corresponding to the Bianchi identities, which have a geometrical meaning rather than dynamical.

An entirely similar situation is present in the case of loop quantum gravity with the called Kodama state \cite{3}; in these cases, instantons pick up a topological phase among available ground states, achieving that the dynamical and topological phases of the theory make contact. The general relationship between the Chern classes and the representation of their ground states by means of the corresponding Chern form is established in \cite{2a, 6}.

In the present case, we have a different situation, since considering the property C) of Section II, and the self-dual conditions, the Lagrangian can be rewritten as the derivative of the ground state functional argument, plus a term that does not vanish as an identity, but represents in general another dynamical sector:
\begin{equation}
\int
  P^{\sigma\alpha\beta} P_{\sigma\alpha\beta}\quad \overrightarrow{_{\cal P=\pm \ast P}}
   \int  \overline{\nabla}_{\lambda}[\varepsilon^{\lambda}{_{\sigma}}P^{\alpha\beta\sigma}H_{\alpha\beta}] -\int H_{\alpha\beta}\widetilde{\nabla}_{\lambda}[\varepsilon^{\lambda}{_{\sigma}}P^{\alpha\beta\sigma}],
\quad \psi_{\rm {bound-state}}= e^{\int[\varepsilon \cdot P \cdot H]};
\end{equation}
where $\widetilde{\nabla}_{\lambda}[\varepsilon^{\lambda}{_{\sigma}}P^{\alpha\beta\sigma}]$ does not vanish as an identity (as in the case discussed just above in relation to the Bianchi identities); rather such a term represents the original equations of motion under the conditions ${\cal P=\pm \ast P}$,
\[
\widetilde{\nabla}_{\lambda}P^{\lambda\alpha\beta}=0
\stackrel{{\cal P=\pm \ast P}}{\longleftrightarrow} \widetilde{\nabla}_{\lambda}[\varepsilon^{\lambda}{_{\sigma}}P^{\alpha\beta\sigma}]=0,
\]
in this manner, instantons in the present context achieve that two dynamical phases of the theory make contact (without picking up a topological phase among available ground states, as opposed to the case discussed in Eq.\ (32)); however, both dynamical phases will have the same bound state. Note that in spite of these crucial differences between (32) and (33), in both cases the ground state is given essentially by the argument of the boundary term.

Furthermore, there is an additional argument that reinforces the above difference in favor to our functionals as genuine quantum states. One reason found in \cite{5} explaining `why' the Chern-Simons state exists is that the associated self-dual conditions define a Lagrangian submanifold {\it N} of the phase space {\it M} of the Yang-Mills theory, and the state is identified with a state annihilated by the operators obtained by quantizing functions that vanish on {\it N}, being hence a trivial state; this fact is proved by showing that the symplectic structure on {\it M} vanishes when restricted to {\it N}. Specifically the symplectic structure of the theory vanishes under the self-dual conditions defining instanton configurations;
the proof depends essentially on the deformation of the self-dual conditions, and on the Fermi statistics for the field deformations understood as one-forms on the phase space. However, in the present case, one can prove that the symplectic structure $\omega$ obtained by deriving the symplectic potential $\theta$ described previously:
\[
\omega= \delta\theta = \int_{\Sigma^{C}} \delta P^{\sigma\alpha\beta} \delta
H_{\alpha\beta} d \Sigma^{C}_{\sigma} \quad \overrightarrow{_{\cal \delta P=\pm \ast \delta P}}\quad \int_{\Sigma^{C}} {\varepsilon^{\sigma}}_{\nu}
\widetilde{\nabla}^{\alpha}   \delta H^{\nu\beta} \delta
H_{\alpha\beta} d \Sigma^{C}_{\sigma}
\]
 does not vanish under the self-duality conditions (22), specifically under the first-order deformations of such conditions and the Fermi statistics for $\delta H$.\\

\noindent {\uno VII. Comparison with the usual gravity propagation on the world-surface}\vspace{1em}

In order to put in perspective the present results concerning with gravity waves propagation governed by the variations of a topological quantity, it is important to make a comparison with the standard ones, which concern also with gravity waves propagation but coming from the quadratic action obtained by expanding the world-surface curvature $R(n)\rightarrow R(n+h)$; for a world-surface fixed, such a quadratic action will lead variationally to linear equations of motion for the deformation field $h$. Considering the expression for the curvature given in \cite{2}, we have the Lagrangian quadratic in $h$:
\begin{equation}
     \Delta R \equiv 2n^{\alpha\beta} h^{\rho}_{\lambda} \overline{\nabla}_{[\rho} \delta \rho^{\lambda}_{\beta]\alpha},
\end{equation}
where $\delta\rho$ was obtained in \cite{1} and is given by
\begin{eqnarray}
     \delta \rho_{\lambda}{^{\mu}}{_{\nu}} \!\! & = & \!\! \overline{\nabla}^{\mu} h_{\nu\lambda} - \overline{\nabla}_{\nu} h^{\mu}_{\lambda} + 2K_{\nu}{^{\rho}}{_{(\lambda}} h^{\mu)}_{\rho} - 2 K^{\mu\rho}{_{(\lambda}} h_{\nu)\rho} \nonumber \\
     \!\! & = & \!\! \widetilde{\nabla}^{\mu} h_{\lambda\nu} - \widetilde{\nabla}_{\nu} h^{\mu}_{\lambda};
\end{eqnarray}
using this expression we can attempt to rewrite $\Delta R$ in terms of the traceless tensor $H_{\mu\nu} \equiv h_{\mu\nu}- \frac{1}{2} n_{\mu\nu}h$, for a direct comparison with the action (1); modulo total derivatives, the result is
\begin{eqnarray}
     \Delta R \!\! & = & \!\! - \widetilde{\nabla}_{\sigma} H_{\alpha\beta} \cdot \widetilde{\nabla}^{\sigma} H^{\alpha\beta} + \widetilde{\nabla}^{\alpha} H_{\alpha\beta} \cdot \widetilde{\nabla}_{\sigma} H^{\sigma\beta} + \widetilde{\nabla}^{\sigma} H_{\alpha\beta} \cdot \widetilde{\nabla}^{\alpha} H^{\beta}_{\sigma} \nonumber \\
     \!\! & & \!\! + (\widetilde{\nabla}^{\alpha} H^{\beta}_{\tau} - \widetilde{\nabla}_{\tau} H^{\alpha\beta}) [K^{\rho\tau}{_{\alpha}} H_{\beta\rho} - K^{\rho}{_{\beta}}{^{\tau}} H_{\alpha\rho} - n^{\tau}_{\beta} K^{\sigma\rho}{_{\alpha}} H_{\sigma\rho} + K^{\tau} H_{\alpha\beta}] \nonumber \\
     \!\! & & \!\! + terms (H,h);
\end{eqnarray}
where $terms(H,h)$ indicate those terms depending on $H$ and the trace $h$; thus, the first difference with the Lagrangian (1) is that $\Delta R$ is not in general a functional of $H$. We do not display explicitly these terms, since we impose now the gauge fixing  $h=0$ on the trace and continue with the comparison, since it may be possible that under appropriate gauge fixing conditions both Lagrangians may be equivalent. The first terms in both Lagrangians are equivalent and will be responsible for the D'Alambertian term in the equations of motion; the second terms are essentially the same, and although they differ by a factor 2, it can be gauged away imposing the Coulomb gauge  (4), further continuing the comparison. However, beyond these terms the differences emerge; the third term in (36) is not present in the Lagrangian (1) and of course it can not be gauged away. Furthermore, the fourth term in (36) depending on $H$ and its $\widetilde{\nabla}$-derivatives is not present in the Lagrangian (1), which is in fact a polynomial in the $\widetilde{\nabla}$-derivatives of $H$. Note that only the penultimate term of the form $nKH$ in the 2-factor of this last term
 can be gauged away under the gauge fixing conditions considered above.

Variationally the Lagrangian (36) leads to the following equations of motion for the deformation field $H_{\alpha\beta}$:
\begin{equation}
 \widetilde{\Box} H^{\alpha\beta} + \frac{1}{2}K_{\tau}\widetilde{\nabla}^{(\alpha}
      H^{\beta)\tau}-\frac{1}{2}K^{\tau(\alpha}{_{\sigma}}[\widetilde{\nabla}_{\tau}H^{\beta)\sigma}+\widetilde{\nabla}^{\beta)}H_{\tau}^{\sigma}]
      +\frac{1}{2}K_{\rho}{^{\sigma(\alpha}}\widetilde{\nabla}_{\sigma}H^{\beta)\rho}
     -\frac{1}{2}H_{\rho\sigma}K^{\alpha\rho\gamma}K^{\beta\sigma}{_{\gamma}} = 0;
     \label{eq-motion-2}
\end{equation}
where the gauge condition (4) as well as the conformal gauge $R=0$ (see section II) have already been considered in order to simplify the expression. For simplicity, a term proportional to the ambient curvature has been omitted considering a flat background spacetime. Additionally from the variations of the Lagrangian (36) the corresponding canonical pair reads $(Q^{\sigma\alpha\beta}=\widetilde{\nabla}^{\sigma} H^{\alpha\beta}-\widetilde{\nabla}^{(\alpha} H^{\beta)\sigma}+\frac{1}{2}(K_{\rho}{^{\sigma(\alpha}}H^{\beta)\rho}+K_{\rho}{^{(\alpha\beta)}}H^{\sigma\rho}-K^{(\alpha}H^{\beta)\sigma}), H_{\alpha\beta})$; thus, since the canonical pairs (commuting functionals on the phase space) are different, the symplectic geometries  of the phase space are evidently distinct; of course, the phase spaces in turn are also different, since the space of solutions of the corresponding equations of motion do not coincide. In relation to the points A), B), and C) at the end of section II, we have now for this case:

A') $Q^{\sigma\alpha\beta}=P^{\sigma\alpha\beta}-\widetilde{\nabla}^{(\alpha} H^{\beta)\sigma}+\frac{1}{2}(K_{\rho}{^{\sigma(\alpha}}H^{\beta)\rho}+K_{\rho}{^{(\alpha\beta)}}H^{\sigma\rho}-K^{(\alpha}H^{\beta)\sigma})$; note that in relation to the canonical commutators,
in the first approach one has $[P,H]=0$, and $[Q,H]\neq 0$; conversely in the approach at hand one has $[Q,H]=0$, and $[P,H]\neq 0$.

B') the canonical momentum $Q$ is not conserved; the $\widetilde{\nabla}$-divergences of $Q$ involve in general derivatives of the second fundamental tensor $K$, leading to the appearance of the third fundamental tensor \cite{2}, and consequently there is no simple relation with the equations of motion.

C') the Lagrangian (36) is not a quadratic functional on the momentum $Q$.

D') there no exists a similar consistent conservation law between metric deformations of the world-surface and their possible sources in this case; the first-order deformation of the Bianchi identities produces naturally a consistent result only for the perturbative treatment obtained from the deformation of the equations of motion treated throughout this paper. In the same sense, one can start just from these deformed Bianchi identities and identify the equations of motion for world-sheet metric deformations satisfying a consistent conservation law when a source is included. The integration of these equations of motion over the world-surface leads to the Lagrangian (1), and in particular the interacting term of the form $H_{\alpha\beta}\overline{S}^{\alpha\beta}$.
However, in the case at hand, the starting point is a Lagrangian quadratic in the field
$H$, and at this level the possible sources $J$ must be considered; for $J$ independent of $H$ a term of the form $H\cdot J$ yields a similar term (the inhomogeneous one) appearing in the first case; additionally a term of the form $H_{\mu\nu}H^{\nu}_{\alpha}J^{\mu\alpha}$, leads to a linear term at the level of equations of motion. Thus, a consistent conservation law can not be obtained in a simple way, and an iterative perturbative treatment must be implemented following for example [8]; more discussion about this issue is given in the concluding remarks.

 E') In the first case, the $n^{\alpha\beta}$-trace of the equations of motion is trivially satisfied due to the tracelessness of $H$. However, such a trace of the motion equations (37) leads to a {\it scalar} reestriction, $K^{\sigma\alpha\beta}[\widetilde{\nabla}_{\sigma}H_{\alpha\beta}
+\widetilde{\nabla}_{\alpha}H_{\sigma\beta}+K_{\rho\alpha\beta}H^{\rho}_{\sigma}]=0$, nonexistent in the first case.

Based on these results, one can develop for the case at hand the Hamiltonian quantization along the same lines; however, the differences at Lagrangian level and at the level of the symplectic structure of the phase space, allow us to see the consequences at quantum level. First, the corresponding energy-momentum tensor defined as the functional derivative of the Lagrangian (36) with respect to the world-surface metric will be different, and in particular the component associated to the {\it Hamiltonian}; considering the appearance of additional constraints, this new scenario will lead to different ground state functionals. The possible instanton field configurations associated with these functionals will be different at both classical and quantum levels. It will be interesting to continue with the comparison at this level, but it is beyond the purposes of this work and will be left for future works.

As we have seen there exist profound differences between both approaches, and a possible coincidence in some classical and/or quantum aspect will not imply
the physical or geometrical equivalence of these wave propagations. For example, one may consider particular world-sheet scenarios, and try to find similarities; the simplest case of a  world line in a two-dimensional ambient space is discarded since the degrees of freedom are restricted for $n\geq 3$ for gravity waves considered in this work[1]; thus, one can consider for example $n=3$ and a cylinder, a sphere, a torus, etc, as world-surfaces for the propagation of gravity waves in both approaches; however the equations of motion show  important differences; in the first case the gravity waves are governed only by the part associated to the D'Alambertian $\widetilde{\Box}$, while the waves
governed by equations (37) are affected, even in these particular scenarios, by additional non-trivial terms. Another world-surface scenario may be the so-called {\it minimal} surfaces for which $K^{\mu}=0$ (the equations of motion of the Dirac-Nambu-Goto action correspond to these surfaces); in this case only a term in the Lagrangian (36), the equations of motion (37), and the momentum $Q$ vanishes, but terms depending on other components of the second fundamental tensor $K$ remain  nontrivial. In fact, both approaches coincide only in the trivial case of a vanishing second fundamental tensor, describing (flat) waves propagating on a flat background scenario.

In spite of these physically meaningful differences, there are general similarities of these perturbative treatments, since both break down the original topological invariance of the Euler characteristic, generating the gravity propagating modes; in this sense, the first paragraph of the concluding remarks on the transition of a phase with no metric to a new phase with such a structure is valid for both approaches.\\

\noindent {\uno VIII. Gravity wave propagation on $S^{2}$}
\vspace{.5em}

We analyze now the gravity wave propagation on simple geometries, starting with $S^{2}$ imbedded in $(R^{3}, \delta)$; thus, with the usual induced metric on $S^{2}$ we can determine the components of the first fundamental tensor $n^{\mu\nu}$ on $S^{2}$:
\begin{eqnarray}
     (n^{\mu\nu}) = \frac{1}{a^{2}}
     \left( \begin{array}{ccc}
     a^{2}-x^{2} & -xy & -xz \\
     -xy & a^{2}-y^{2} & -yz \\
     -xz & -yz & x^{2}+y^{2}
     \end{array} \right) ;
\label{matrix-n-S2}
\end{eqnarray}
where $x^{2}+y^{2}+z^{2}=a^{2}$. The relevant differential operators in this formulation are given by $\overline{\partial}_{\mu} = n^{\nu}_{\mu} \partial_{\nu}$, explicitly
\begin{eqnarray}
     \overline{\partial}_{x} \!\! & = & \!\! \frac{1}{a^{2}} (2a^{2}-2x^{2}-y^{2}) \partial_{x} + \frac{1}{a^{2}} \frac{x}{y} (a^{2}-2y^{2}-x^{2}) \partial_{y}, \nonumber \\
     \overline{\partial}_{y} \!\! & = & \!\! \frac{1}{a^{2}} \frac{y}{x} (a^{2}-2x^{2}-y^{2}) \partial_{x} +  \frac{1}{a^{2}} (2a^{2}-2y^{2}-x^{2}) \partial_{y}, \nonumber \\
     \overline{\partial}_{z} \!\! & = & \!\! - \frac{1}{a^{2}} \frac{z}{x} (2x^{2}+y^{2}) \partial_{x} - \frac{1}{a^{2}} \frac{z}{y} (2y^{2}+x^{2}) \partial_{y};
\label{operator-S2}
\end{eqnarray}
where we have considered that $z=z(x,y,a)$ from the constraint that defines $S^{2}$. Hence, the ``D' Alambertian" operator is given by
\begin{equation}
     \overline{\partial}^{\mu} \overline{\partial}_{\mu} = \frac{x^{2}+z^{2}}{z^{2}} \overline{\partial}^{2}_{x} + \frac{xy}{z^{2}} \overline{\partial}_{x} \overline{\partial}_{y} + \frac{1}{a^{2}z^{2}x} [ x^{4}-y^{4}+a^{2}(2x^{2}+y^{2})] \overline{\partial}_{x} + (x \leftrightarrow y),
\label{D'Alambertian-S2}
\end{equation}
where $(x \leftrightarrow y)$ represents a similar term to the first one but interchanging $x$ with $y$. Furthermore, considering Eqs. (\ref{matrix-n-S2}) and (\ref{operator-S2}) the components of the second fundamental tensor are given by
\begin{eqnarray}
     K_{11}{^{i}} \!\! & = & \!\! \frac{1}{a^{4}} (2x^{2}+y^{2}-2a^{2})x^{i}, \qquad K_{12}{^{i}} = \frac{1}{a^{4}}\frac{y}{x} (2x^{2}+y^{2}-a^{2})x^{i}, \qquad K_{13}{^{i}} = \frac{1}{a^{4}} \frac{z}{x} (2x^{2}+y^{2})x^{i}, \nonumber \\
     K_{23}{^{i}} \!\! & = & \!\! \frac{1}{a^{4}} \frac{z}{y} (2y^{2}+x^{2})x^{i}, \qquad K_{22}{^{i}} = \frac{1}{a^{4}} (2y^{2}+x^{2}-2a^{2})x^{i}, \qquad K_{33}{^{i}} = -\frac{3}{a^{4}} (x^{2}+y^{2})x^{i}, \nonumber \\
     K^{i} \!\! & = & \!\! n^{\mu\nu} K_{\mu\nu}{^{i}} = -\frac{K}{a^{6}}x^{i}, \qquad K = 4a^{4}+x^{2}(a^{2}-x^{2})+y^{2}(y^{2}-a^{2});
\label{K}
\end{eqnarray}
with these elements we can find an explicit form for the equations of motion (\ref{eq-motion}), by considering the expression for the operator $\widetilde{\Box}$ given in \cite{1} (see Eqs. (71) in that reference):
\begin{equation}
     \overline{\partial}^{\lambda} \overline{\partial}_{\lambda} H_{\mu\nu} + 2 K_{\sigma}K^{\lambda}{_{(\mu}}{^{\sigma}} H_{\nu )\lambda} =0;
\label{exp-eq-mov}
\end{equation}
where we have considered that $R=0$ according to the conformal gauge, and that the background space is flat;
the operator $\overline{\partial}^{\lambda} \overline{\partial}_{\lambda}$ is given explicitly in Eq. (\ref{D'Alambertian-S2}). The term quadratic in $K$ plays the role of an {\it effective} potential $V$ on the wave propagation, explicitly reads
\begin{equation}
     (V^{\lambda}{_{\mu}}) \equiv (K_{\sigma} K^{\lambda}{_{\mu}}{^{\sigma}}) = \frac{K}{a^{8}}
     \left( \begin{array}{ccc}
     2a^{2}-2x^{2}-y^{2} & \frac{y}{x} (a^{2}-2x^{2}-y^{2}) & -\frac{z}{x} (2x^{2}+y^{2}) \\
                         & 2a^{2}-2y^{2}-x^{2} & -\frac{z}{y} (2y^{2}+x^{2}) \\
                         &                     & 3(x^{2}+y^{2})
     \end{array} \right) ;
\label{potential-S2}
\end{equation}
where the absent components in the matrix can be obtained by considering that the potential is symmetric;
this potential couples some of the components of the field $H_{\mu\nu}$, for example the component $"11"$ of Eq.
(\ref{exp-eq-mov}) $(\overline{\Box}+V^{1}{_{1}})H_{11} + V^{2}{_{1}}H_{12} + V^{3}{_{1}}H_{13}=0$. However, due to  the constraints (4), only two components of $H$ are independent; thus these propagating modes have two degrees of freedom on $S^{2}$ imbedded in $R^{3}$ (see Section 6 in (\cite{1})).

On the other hand, one of the differences between the equations (\ref{exp-eq-mov}) and
(\ref{eq-motion-2}) is  the form of the {\it effective} potential, which needs a
shift respect to the expression (\ref{potential-S2}); explicitly such a potential can be written
 as $U_{\mu\nu}^{\sigma\rho}\equiv 2 \delta^{\sigma}_{(\mu}V_{\nu)}^{\rho}-\frac{1}{2} K_{\mu}{^{\sigma\gamma}}K_{\nu}{^{\rho}}{_{\gamma}}$, which is again symmetric and quadratic
  in $K$. The shift can be determined explicitly as a function of $(x,y,a)$ using the
  expressions (\ref{K}) for the second fundamental tensor; the coupling term will be then of the form
 $U_{\mu\nu}^{\sigma\rho}H_{\sigma\rho}$.\\

 \noindent {\uno IX. Gravity wave propagation on $T^{2}$}
\vspace{.5em}

Following the same geometric considerations on the imbedding and induced metric in $(R^{3}, \delta)$ of the previous case, the first fundamental tensor on $T^{2}$ is given by
\begin{eqnarray}
     (\eta^{\mu\nu}) =
     \left( \begin{array}{ccc}
     \frac{x^{2}z^{2}+a^{2}y^{2}}{a^{2}(x^{2}+y^{2})} & -\frac{1}{2} \sin2\mu \cos^{2}\nu & -\frac{1}{2} \sin2\nu \cos\mu \\
     \ldots              & \frac{y^{2}z^{2}+a^{2}x^{2}}{a^{2}(x^{2}+y^{2})} & -\frac{1}{2} \sin2\nu \sin\mu \\
     \ldots              & \ldots                    & \frac{a^{2}-z^{2}}{a^{2}}
     \end{array} \right) ;
\label{matrix-n-T2}
\end{eqnarray}
where $(c-\sqrt{x^{2}+y^{2}})^{2}+z^{2}=a^{2}$ on $T^{2}$. In this expression for the fundamental tensor the components out of the diagonal are expressed in terms of the pair $(\mu,\nu)$ for simplicity, but they must be understood actually in terms of the background coordinates $(x,y,z)$, considering the (inverse) imbedding functions $x=(c+a\cos \nu)\cos \mu, y=(c+a\cos \nu)\sin \mu$, and $ z= a\sin \nu$. The absent components below the diagonal have been omitted for simplicity and can be obtained by considering that the tensor $\eta^{\mu\nu}$ is symmetric. Additionally, we consider that $z=z(x,y)$ from the relation described above. The relevant operators are given in this case by:
\begin{eqnarray}
     \overline{\partial}_{x} \!\! & = & \!\! \frac{1}{a^{2}} \frac{(2x^{2}+y^{2})z^{2}+a^{2}y^{2}}{x^{2}+y^{2}} \partial_{x} + \frac{1}{a^{2}} \frac{x}{y} \frac{(2y^{2}+x^{2})z^{2}-a^{2}y^{2}}{x^{2}+y^{2}} \partial_{y}, \nonumber \\
     \overline{\partial}_{y} \!\! & = & \!\! \frac{1}{a^{2}} \frac{y}{x} \frac{(2x^{2}+y^{2})z^{2}-a^{2}x^{2}}{x^{2}+y^{2}} \partial_{x} + \frac{1}{a^{2}}  \frac{(2y^{2}+x^{2})z^{2}+a^{2}x^{2}}{x^{2}+y^{2}} \partial_{y}, \nonumber \\
     \overline{\partial}_{z} \!\! & = & \!\! n^{1}_{3}\frac{2x^{2}+y^{2}}{x^{2}} \partial_{x} + n^{2}_{3}\frac{2y^{2}+x^{2}}{y^{2}}\partial_{y} = \frac{a^{2}}{z^{2}} (n^{1}_{3}\overline{\partial}_{x} + n^{2}_{3}\overline{\partial}_{y});
\label{operator-T2}
\end{eqnarray}
and the D'Alambertian operator takes the form
\begin{equation}
     \overline{\partial}^{\mu}\overline{\partial}_{\mu} = \frac{y^{2}z^{2}+a^{2}x^{2}}{x^{2}+y^{2}} \frac{1}{z^{2}} \overline{\partial}^{2}_{x} + \frac{xy}{z^{2}} \frac{a^{2}-z^{2}}{x^{2}+y^{2}} \overline{\partial}_{x}\overline{\partial}_{y} + \frac{a^{2}}{z^{2}} [n^{1}_{3} \overline{\partial}_{x} (\frac{a^{2}}{z^{2}} n^{1}_{3}) + n^{2}_{3} \overline{\partial}_{y} (\frac{a^{2}}{z^{2}}n^{1}_{3})]\overline{\partial}_{x} + (x\leftrightarrow y).
\label{DAlambertian-T2}
\end{equation}
Eq. (\ref{exp-eq-mov}) as stands is valid under the only assumption of a flat background and the use of the conformal gauge; the difference with the case of a $S^{2}$ world-sheet scenario is that the D'Alambertian operator is given now by Eq. (\ref{DAlambertian-T2}), and the effective potential $V^{\lambda}{_{\mu}}=K_{\sigma} K^{\lambda}{_{\mu}}{^{\sigma}}$ can be constructed explicitly as a function on $(x,y,z,c,a)$ from the expression for the first fundamental tensor (\ref{matrix-n-T2}) and the operators (\ref{operator-T2}). For the second approach describing the wave propagation, we shall have similarly a new potential with the same dependence on $V^{\lambda}{_{\mu}}$ and the shift corresponding to a quadratic term in $K$ considered in the $S^{2}$ case. The explicit expressions are very involved and have been omitted for simplicity; we shall consider such expressions in future treatments. However, the general features in relation with the coupling between the components of the metric deformations due to the presence of the potential, and the counting of degrees of freedom considered in the $S^{2}$ case are valid also on $T^{2}$.\\

\noindent {\uno X. Concluding remarks}\vspace{1em}

In the two previous sections we have applied our general results to two compact world-surfaces without boundaries, since the starting point in \cite{1} was implicitly
the expression of the Euler characteristic in terms of the scalar world-surface curvature only. Then, the results can not be applied directly on compact world-surfaces such
as a cylinder, due to the presence of the boundaries ($S^{1}$), which have contributions
to the Euler characteristic; however our results can be generalized in straightforward form.

It is convenient to sum up the results from references \cite{1,7} in order to put in perspective the new ones. The two-spin field theory considered here was obtained by breaking down the topological invariance of the action that emerges naturally in string theory, the Euler characteristic of the world-surface \cite{1}; this topological phase of the theory (a phase with no metric) is characterized by certain Wilson loops defined on the world-surface, and that in particular can be reduced to a Wilson loop along the spatial configuration of the (closed) string \cite{7}. When the topological invariance is broken perturbatively, then a spin-two field emerges, generating then a phase with a metric \cite{1}.
This phase is described by the action (1), which shows invariance under diffeomorphisms as a remanent symmetry of the original topological invariance of the unbroken phase described by the Wilson loops. Therefore, the spin-2 field corresponds to the physical spectrum of excitations around the unbroken phase, and the ground state of these excitations are described by the wave functionals found here. Such functionals can be considered as the string versions of the Chern-Simons states found in the context of Yang-Mills theory and quantum gravity.

Nowadays there exists a controversy with respect to the viability of the Chern-Simons as a quantum state; for example in \cite{5} it is claimed that due to the unnormalizability, lack of CPT-invariance and additional unwanted properties, such a state does not correspond to a sensible physical theory. However, in the context of quantum gravity, the conclusion in the second of Ref. \cite{3}  is that although the unnormalizability is inevitable in the linearized Kodama state (in which general covariance is broken), the argument can not be extended to the full state which is a generally covariant state, and there is not a definitive conclusion to respect. However, beyond the Kodama state  in loop quantum  gravity there exist general solutions where they are physically aceptable as quantum states of the gravitational field, the so-called spin network states.  Hence, in the worse scenario, the Kodama state is just a semiclassical approximation to the quantum state of gravity.
In the present context,
we need to explore if the ground states constructed are (or not) normalizable under an appropriate inner product. In fact, the  inner product can be obtained by means the  Faddeev-Jackiw or Hamilton-Jacobi  formulations just like for unphysical and physical theories  is obtained in  \cite{9a, 10}, and it is mandatory work in this direction; for example, the {\it naive} ``inner product"
\[
<\psi_{0}\mid\psi_{0}>=\int d H \mid\psi_{0}(H)\mid^{2}=\int d H e^{2\int[\varepsilon \cdot P \cdot H]},
\]
is not finite, since there exist unbounded directions in the field space; more work is clearly needed.

We discuss briefly some technical aspects that will remain to be worked out.
One can verify that $T^{\mu\nu}$ is not world-surface ($
\widetilde{\nabla}-$)conserved; although this property may be considered as unwanted,
actually is expected under general perturbative considerations, since if $T^{\mu\nu}$
is considered as a source for the field $H^{\mu\nu}$ in a second order perturbative treatment, then the invariance under diffeomorphisms of the corresponding self-interacting term of the form
$T^{\mu\nu}H_{\mu\nu}$, requires that $T^{\mu\nu}$  is not separately world-surface conserved, because
$T^{\mu\nu}$ is itself affected by the action of diffeomorphisms. This issue will be extended in a complete second order treatment in progress (see for example \cite{8} for a recent criticism on perturbative treatments for graviton field theory). The appearance of non-linear effects in this second-order treatment will affect the possible (un)normalizability of the wave functionals constructed.

On the other hand,
instantons are associated with finite moduli spaces, which correspond to the space of solutions of the equations of motion divided by the volume of the symmetry group. There is no guarantee {\it a priori} that the moduli space is of finite dimension; however, it is possible that with the self-duality or anti-self-duality conditions for instantons, complemented with the gauge condition (4), an elliptic complex associated with a finite moduli volume may be constructed. In this case the partition function constructed in \cite{1} for the spin-2 field theory may be expressed only in terms of instantons, or anti-instantons contributions. This will be a problem for the future.
Another problem for the future is on the possibility of obtaining, in an appropriate limit, classical general relativity or quantum spin-2 field theory from the present results. Additionally it will be interesting to explore some cosmological applications, to find explicit solutions for the self-dual, and anti-self-dual conditions.\\

\begin{center}
{\uno ACKNOWLEDGMENTS}
\end{center}
This work was supported by VIEP-BUAP and the Sistema Nacional de Investigadores
 (M\'{e}xico).

\end{document}